\documentclass{PoS}

\usepackage{amsmath}
\usepackage{amsfonts}
\usepackage{amssymb}
\usepackage{amsthm}
\usepackage{slashed}
\usepackage{subfig}

\newcommand{\ssm}{AFM}

\newcommand{\preprintline}{\newline
\vskip -5.2cm
\rightline{\parbox{4cm}{\large\rm IFT-UAM/CSIC-10-58 \\ FT-UAM-10-29 }}
\vspace{4.2cm}}

\title{Probing the Yang-Mills vacuum with adjoint zero-modes \preprintline}

\ShortTitle{Probing the Yang-Mills vacuum with adjoint zero-modes}

\author{Margarita Garc\'ia P\'erez \\%
        Instituto de F\'isica Te\'orica UAM-CSIC \\
        Universidad Aut\'onoma de Madrid, Cantoblanco, 28049 Madrid, Spain\\
        E-mail: \email{margarita.garcia@uam.es}}

\author{Antonio Gonz\'alez-Arroyo \\%
        Departamento de F\'isica Te\'orica and Instituto de F\'isica Te\'orica UAM-CSIC \\
        Universidad Aut\'onoma de Madrid, Cantoblanco, 28049 Madrid, Spain\\
        E-mail: \email{antonio.gonzalez-arroyo@uam.es}}

\author{\speaker{Alfonso Sastre} \\%
        Departamento de F\'isica Te\'orica and Instituto de F\'isica Te\'orica UAM-CSIC \\
        Universidad Aut\'onoma de Madrid, Cantoblanco, 28049 Madrid, Spain\\
        E-mail: \email{alfonso.sastre@uam.es}}

\abstract{For the non-perturbative analysis of the topological content of Yang-Mills theories, it is essential
to disentangle long-range structures from short-range fluctuations. Some time ago, 
one of us proposed to use the adjoint modes of the Dirac operator for this purpose. In this talk we 
analyse an implementation of this idea that associates two Weyl fermionic modes in the adjoint 
representation to every gauge field configuration. The densities of these modes provide a 
filtered image of the self-dual and anti self-dual parts of the gauge action density. We present successful 
tests on the performance of this proposal on a set of initial gauge field configurations.}

\FullConference{The XXVIII International Symposium on Lattice Field Theory\\
                 June 14-19,2010\\
                 Villasimius, Sardinia Italy}

\begin{document}

%

\section{Introduction}

In the past years several  authors have devoted some effort to the analysis of 
the topological structures present  in  pure Yang-Mills theory on the lattice.
Apart from global information such as the topological susceptibility,
it is interesting to extract  local one, such as  the instanton size distribution.
In some cases one might be able to check certain ideas and proposals 
relating these topological structures  to chiral
symmetry breaking and confinement. In this program the battlehorse is 
the  roughness of lattice Monte-Carlo configurations, having its origin 
in divergent  ultraviolet fluctuations. To tame this noisy background
several ideas have been proposed.  In particular,  
cooling  and  smearing  algorithms produce smoother
configurations which are assumed to preserve the long-range structure
of the original one. However, these methods have been
criticised since they may produce some   distortions  on the
 shape distribution of local structures.  
A proposed  alternative is to use filtering methods based on the Dirac operator or other
differential operators \cite{Bruckmann2005} (see for example \cite{Ilgenfritz2007} and references therein).
For the case of the Dirac operator, the main idea is the relation between fermions and topology given by
the Atiyah-Singer index theorem and the correlation between the gauge action
density and the local density of the eigenstates of the Dirac
operator. In practise, however, these methods are not totally free of ambiguities. 
For example, when fermions
in the fundamental representation are used to reconstruct the topological charge density
the ability to reproduce the topological structure depends, in a rather strong way, on the number
of modes included in the reconstruction. In any case, even for
noiseless configurations the shape of the filtered density does  not
coincide with the corresponding action density. 

In this talk we analyse an alternative proposal,  presented  in \cite{Gonzalez-Arroyo2006}, 
based on the use of the adjoint representation of the Dirac  operator.
The advantage of this method is that it is based upon a single mode
and gives a perfect match for classical solutions of the equations of
motion. In the next section we will give details of the method and its
lattice implementation. In Section 3 we will test its performance by 
analysing its filtering capacity for a series of initial gauge field
configuration.

\section{Description of the  method}

For every gauge field configuration we will construct two  associated  classical Weyl 
fermionic fields in the adjoint representation, $\psi_{\pm}(x)$,
whose densities, $||\psi_{\pm}(x)||^2$, correspond to a filtered image of the 
self-dual and anti self-dual parts of the initial gauge field action
density. The Weyl spinorial fields have components  $\psi_{\alpha\, 
\pm}^a(x)$ where $\alpha$ is a 2-component spinor index, and $a$ is
the colour index taking $N^2-1$ values.  From now on we will refer to
the fields $\psi_{\pm}$ as {\em adjoint filtering
modes} or \ssm.

An important property that any filtering method should satisfy is that
for smooth configurations the procedure must reproduce the classical structures without
distortion. The main idea behind the method introduced  in
Ref.~\cite{Gonzalez-Arroyo2006} is that for configurations that are solutions of the classical equations 
of motion (classical solutions)  there is an 
optimal choice of the spinorial field in the adjoint  representation that reproduces 
exactly the shape of  the (anti)self-dual part of the gauge action density. 
This is obtained choosing the \ssm, $\psi_{\pm}$, as the chiral components of the
supersymmetric zero-mode of the Dirac operator, defined by:
\begin{equation}
 \psi = \frac{1}{8}F_{\mu\nu}\left[\gamma_\mu,\gamma_\nu\right] V\,,
\end{equation}
where $V$ is an arbitrary constant spinor. 
This mode  has three interesting properties:\\

\vspace*{-0.20cm}

\indent 1. \ It is a zero-mode: $\slashed{D}\psi = 0$.\\
\indent 2. \ It satisfies the reality condition,
$\textrm{Im}(\psi_{1\, \pm}^{a}(x)) = 0 \quad  \forall a,x$.\\
\indent 3. \ Its density is exactly proportional to the (anti)self-dual part of the gauge action density.\\

\vspace*{-0.20cm}

\noindent Thus, for a classical solution, if we select a mode
satisfying properties  $1$ and $2$, then by virtue of  the third property, 
we obtain an optimal image of the field density structure.

For a general gauge field configuration, our method  consists in
finding the Weyl-spinor  fields in the adjoint representation that
satisfy properties $1$ and $2$ as much as possible. 
Different versions of the method follow from the relative importance
given to both properties. Previous
tests~\cite{Gonzalez-Arroyo2006,latproc}  were done by looking first at
the subspace of low-lying eigenmodes of the adjoint Dirac equation and
then selecting the combination which best satisfies the reality
condition as the \ssm\ mode.

Here we will use an alternative more elegant proposal, also presented in 
Ref.~\cite{Gonzalez-Arroyo2006}, in which the reality condition
(property 2) is imposed exactly. The other condition is implemented by 
requiring that the Weyl spinor is an eigenvector of lowest eigenvalue
of $-\slashed{D}^2$. For a left-handed (positive chirality) spinor this becomes $-D\bar{D}$,
where $D\equiv D_\mu \sigma_\mu$ ($\bar{D}\equiv D_\mu\bar{\sigma}_\mu$) 
are the Weyl operators. Actually, in the adjoint
representation all eigenvalues are doubly degenerate due to euclidean
CP invariance. Then, using any eigenvector $\psi$ and its charge  conjugate
$\psi_C=\sigma_2 \psi^*$ one can form a colour vector of quaternionic
matrices  
\begin{equation}
(\psi, \psi_C)= \Psi_{\pm}^{\mu}(x)  \sigma_{\mu}\,,
\end{equation}
where $\sigma_\mu=(\mathbf{I},-i \vec{\tau})$, 
$\bar{\sigma}_\mu=(\mathbf{I},i \vec{\tau})$ and $\tau_i$ are the
Pauli matrices.  The eigenvalue equation becomes an equation acting 
on quaternionic matrix fields, 
and the reality condition amounts to $\Psi_{\pm}^{ 0 }(x)=0$. 

In summary,  the positive chirality \ssm\ mode is defined by
the eigenvalue condition:
\begin{equation}
-D\bar{D}  \Psi_{+}^{i }(x) \sigma_i= \lambda
\Psi_{+}^{ i }(x) \sigma_i\, ,
\end{equation}
or equivalently 
\begin{equation}
O^+_{ij} \Psi_{+}^{ j }(x) =\lambda \Psi_{+}^{ i }(x)\, ,\ \ 
{\rm where }\ \ \ \ \  
O^+_{jk} = -D_\mu D_\nu \eta_{\alpha}^{\mu\nu}\bar{\eta}_j^{\alpha k}\,,
\end{equation}
and $\eta$ and $\bar{\eta}$ are the `t Hooft symbols.
The last expression can be expanded and we get
\begin{equation}
 O^+_{ij} = -\delta_{ij}D_\mu^2 - \epsilon_{ijk}\eta_{j}^{\mu\nu}D_\mu
 D_\nu =
 -\delta_{ij}D_\mu^2 + i\epsilon_{ijk}(E_k + B_k)\,.
 \label{eq:op_susyp}
 \end{equation}
Notice, that the $O^+$ operator is a positive definite, real,
symmetric operator, so that its eigenvalues are positive real numbers. 
For gauge configurations which are  solutions of the classical
equations of motion the lowest eigenvalue vanishes and generically is
non-degenerate. The corresponding eigenvector is the supersymmetric
zero mode $\Psi_{+}^{ i }(x) \propto  (E_i+B_i) $. 
 
For arbitrary gauge configurations the minimum eigenvalue of $O^+$ is
non-zero. Its corresponding eigenvector  is, by definition, the \ssm\ 
mode. Its density provides the filtered version of the self-dual part
of the action density. The usefulness of the method depends on its
capacity to eliminate high frequency noise without altering the shape
of smooth structures. In this talk we will present the  results of our
tests done on both smooth and rough configurations.

The previous formulae  can be repeated for the negative chirality mode, which
provides a filtered version of the anti-self-dual part of the action
density. The corresponding operator is now replaced by 
\begin{equation}
 O^-_{ij} = -\delta_{ij}D_\mu^2 - i\epsilon_{ijk}(E_k - B_k)\,.  
\label{eq:op_susym}
\end{equation}

In order to obtain a lattice implementation of the filtering procedure
it is more convenient to work with the overlap Dirac operator and 
construct the hermitian, positive definite matrices 
$H^2_\pm = P_{\pm}(\gamma_5 D_{ov})^2 P_{\pm}$, where $P_\pm$ are the
projectors onto the positive and negative chirality modes. The
two-fold degeneracy associated with CP invariance also holds on the
lattice, so that the operators can be taken to act on quaternionic
vectors. Imposing the reality conditions leads to the  
\begin{equation}
 O^{\pm}_L = P_0 H^2_{\pm} P_0\,,
 \end{equation}
where the action of $P_0$ on a quaternionic field is given by 
 $P_0\Psi = \Psi - \frac{1}{2}\sigma_0Tr(\sigma_0\Psi)$.
 
The \ssm\ modes are the eigenvectors of lowest eigenvalue of $O^{\pm}_L$ 
and can be obtained by a conjugate gradient  algorithm. Details of the
technique used  to compute the overlap Dirac operator and other
numerical aspects can be seen in Ref.~\cite{alfonsothesis}

\section{Testing the filtering method}
\label{sc:test}
In this section, we present the results of our tests of the filtering
procedure when applied to various lattice configurations. First, we
apply it to an instanton configuration. Being a classical solution the
method should work well, but it will allow us to quantify 
the finite volume and discretisation effects. Next we apply it to a
series of  instanton-anti-instanton  (IA) pairs with varying
separations. This is intended to monitor possible distortions and
problems which could occur when applied to smooth configurations which
are not classical solutions of the equations of motion. Finally, we
will go back to the single instanton case and add stochastic noise to
it. The filtering method should be able to reduce this noise and
produce a neat image of the underlying smooth configuration.

\subsection{Results for Classical solutions}
\label{sc:ti}
The first test is done over a set of smooth, $Q\!=\!1$, SU(2) instanton
configurations generated by cooling. In this case, the $O^+$ operator should have one 
zero-mode corresponding exactly to the supersymmetric zero-mode. 
As will be discussed below, discretisation and finite volume effects may shift the corresponding eigenvalue 
from zero.  Still, we observe that the density of the lowest eigenvector of the 
$O^+$ operator reproduces to an excellent degree the instanton action density. 
Fitting both the action and the \ssm\ mode densities to the continuum formula, we extract instanton 
positions and sizes that differ at most in $\Delta X= 0.01\,a$, and in 
$\Delta \rho = 0.05\, \rho$, respectively.

In contrast to the situation in the continuum, 
the lowest eigenvalue of $O^{\pm}_L$ is different from zero. 
This is due to discretisation errors and finite volume effects. 
The latter arise for the case of periodic boundary conditions (PBC),
because $Q\!=\!1$ classical solutions do not exist on a periodic 
torus \cite{Braam1989}. This is, however, not the case if twisted
boundary conditions (TBC) are used instead. 

\begin{figure}
 \centering{
 \subfloat[]{\label{fig:inst_lowest}\includegraphics[angle=-90,scale=0.25]{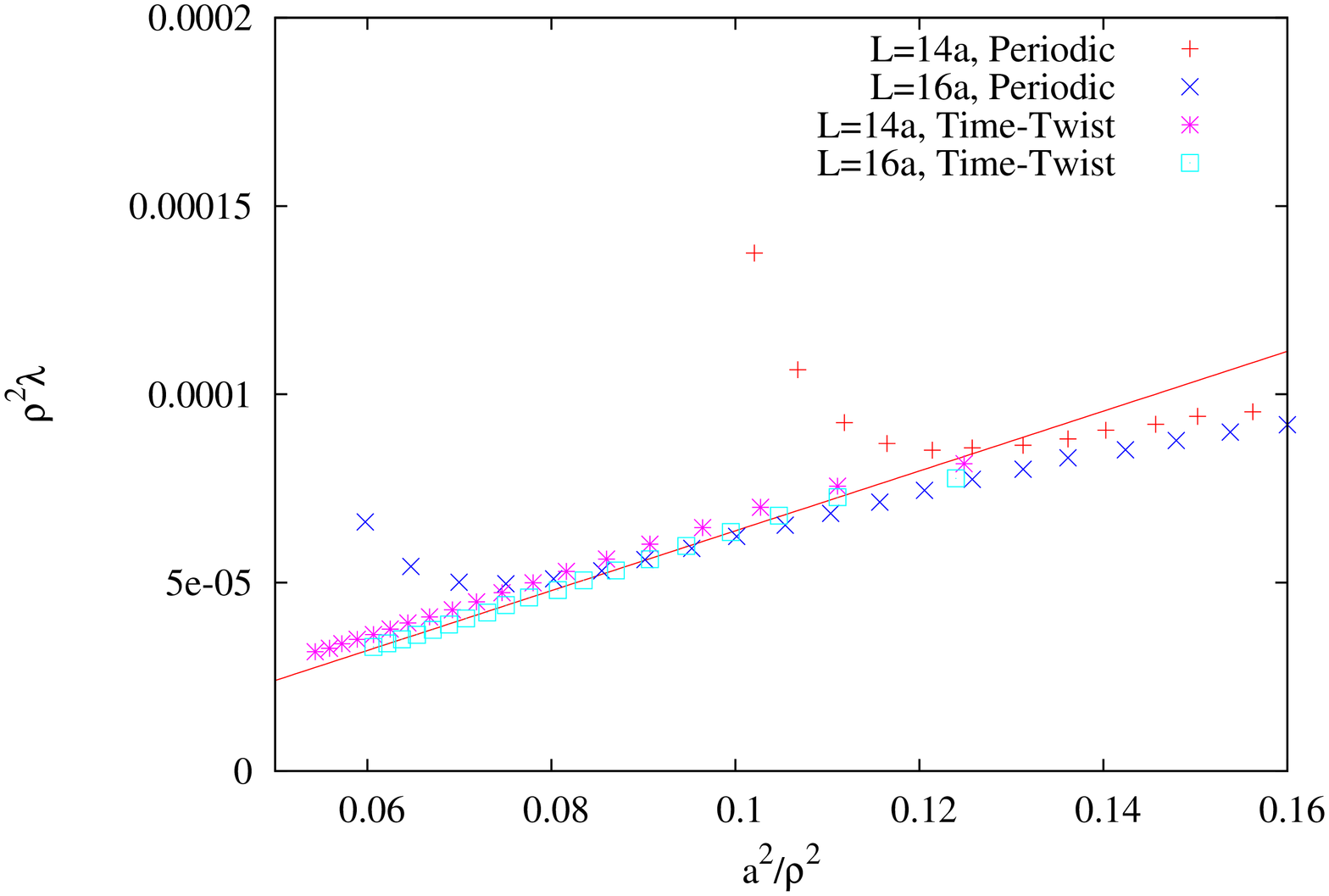}}
 \subfloat[]{\label{fig:ia_ssd}\includegraphics[scale=0.25,angle=-90]{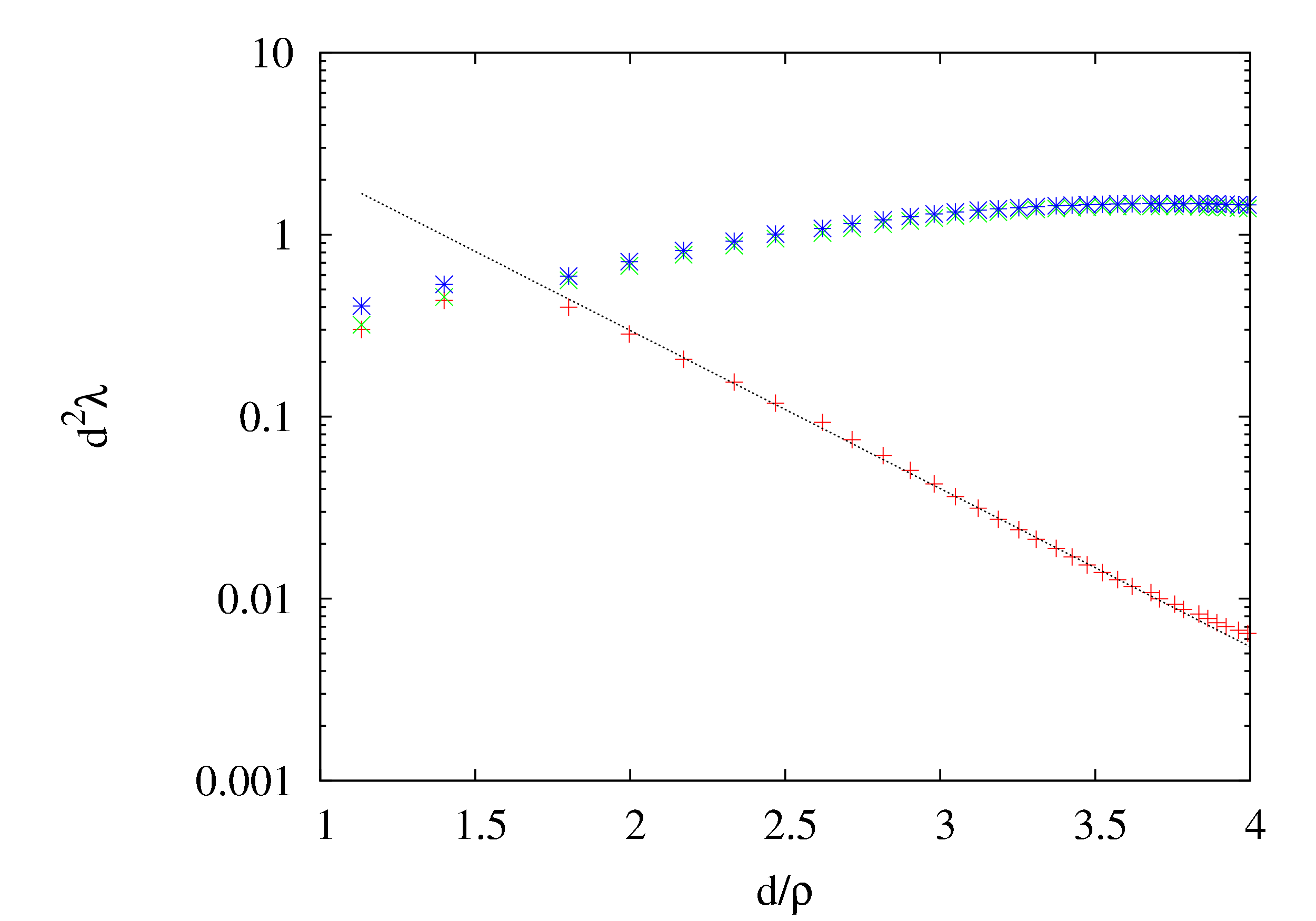}}}
  \caption{(a) The lowest eigenvalue of the $O^+$ operator, on a size $\rho$ instanton 
background, is displayed as a function of 
 $a^2/\rho^2$ for two different lattice sizes and boundary conditions. 
 The finite-volume effects 
 become sizable for periodic BC and $L < 4 \rho$. They are negligible for
 time-twisted BC ($k\!=\!(1,1,1),\;\; m\!=\!(0,0,0)$).
(b) The three lowest eigenvalues of the $O^+$ 
operator, on an IA pair background,  are displayed in terms
 of the IA distance ($d$). Level crossing appears at $d \sim 1.8 \, \rho$.}
\end{figure}

To explore both effects we have generated a large set of 
SU(2) instanton configurations with varying sizes  for both PBC and TBC.
Fig.~\ref{fig:inst_lowest} displays the lowest eigenvalue ($\lambda_1$) of the $O^+$ 
operator as a function 
of the inverse instanton size squared $a^2/\rho^2$.
As expected, $\lambda_1$ approaches zero in the continuum limit as $a^2/\rho^4$, with a coefficient of 
approximately $8\cdot 10^{-4}$.

In what concerns finite-volume effects, a clear distinction can be observed  between periodic and twisted
boundary conditions, as expected.
In the twisted case, classical solutions exist for all torus sizes and
we observe no corrections to the lowest  eigenvalue other that those associated to discretisation effects.
For periodic BC, however, we observe a large deviation from the expected behaviour for
$L \lesssim 4 \rho$. 

\subsection{Smooth non-classical configurations}

\begin{figure}
 \centering{
 \includegraphics[scale=0.25,angle=270]{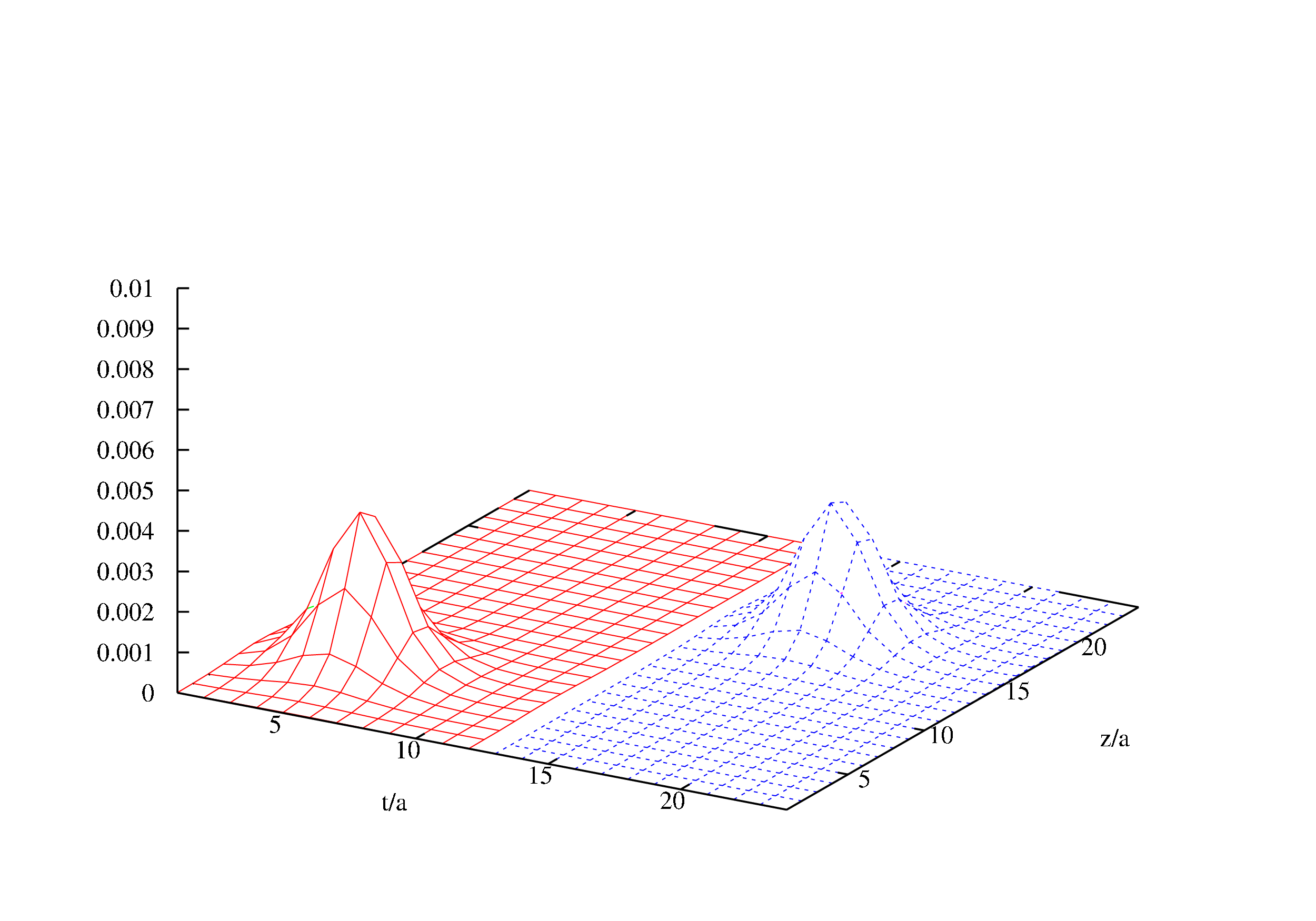} 
 \includegraphics[scale=0.25,angle=270]{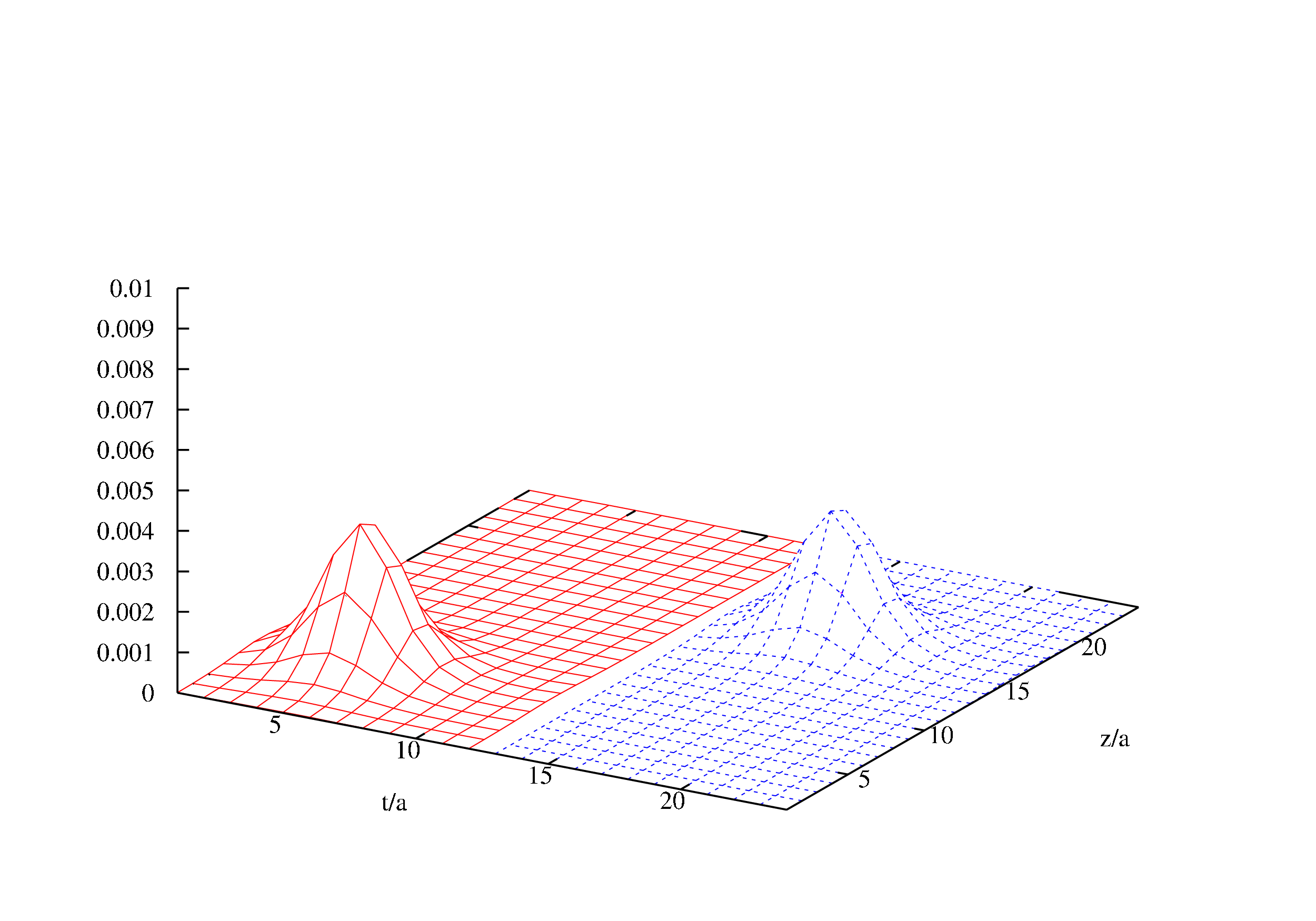}}
\caption{For an instanton-anti-instanton pair, at separation $d\!=\! 3.5 \, \rho$, we 
display
2-d slices of: (Left) the self-dual (red) and the anti-self-dual (blue) parts of the action density. 
(Right) the \ssm\ density  corresponding to the positive (red) and the negative (blue) 
chiralities.}
\label{fig_ia30}
\end{figure}

According to our proposal, even for non-classical configurations the lowest eigenvector 
of the $O^+$ ($O^-$) operator should provide the filtered action densities of the self-dual 
(anti-self-dual) part of the gauge field. Since only for classical solutions there is an exact 
 supersymmetric zero-mode,  it is not guaranteed that 
the filtered image of non-classical configurations has no distortions. In order to test this, we 
have analysed a set of instanton-anti-instanton (IA) configurations with varying separation.    
The set is generated by cooling an initial  well-separated pair and monitoring the different stages
of the IA annihilation process. Fig.~\ref{fig_ia30} displays one snapshot corresponding to IA distance
$d = 3.5 \, \rho$. We compare the \ssm\ densities with the self-dual and anti-self-dual parts of the 
action density. 
The agreement is excellent and, as expected,  the $O^\pm$ lowest eigenvectors are only sensitive 
to objects with the appropriate chirality.
In order to obtain a more quantitative comparison, we extract 
the size parameter of the (anti-) instanton by fitting both the \ssm\ and action densities. Both determinations 
differ  by 5\% at most as long  as the IA separation is larger than $2 \rho$. 

Although the Dirac operator does not have zero-modes on IA backgrounds,
we expect the lowest eigenvalue in each chiral sector to go to zero as the I-A separation is 
increased. To show that this is indeed the case, we display in Fig.~\ref{fig:ia_ssd} the three lowest 
eigenvalues of the $O^+$ operator versus the IA separation.
The smallest one is well described by an  exponentially decreasing
function  of the   IA separation as $d^2 \lambda_1  = 16 \, \exp (-2d/\rho)$.
For distances $d\sim 2 \rho$ we observe a level crossing in the spectrum of the $O^{\pm}$
operators. For even smaller separations the lowest mode  no longer
reproduces the IA gauge action density. Hence, we conclude  that the $O^{\pm}$ operators can identify the 
components of the IA pair as long  as  $d > 2 \rho$. 

\subsection{Configurations with stochastic noise}
\label{sc:heat}

\begin{figure}
\centering{
\includegraphics[scale=0.25,angle=270]{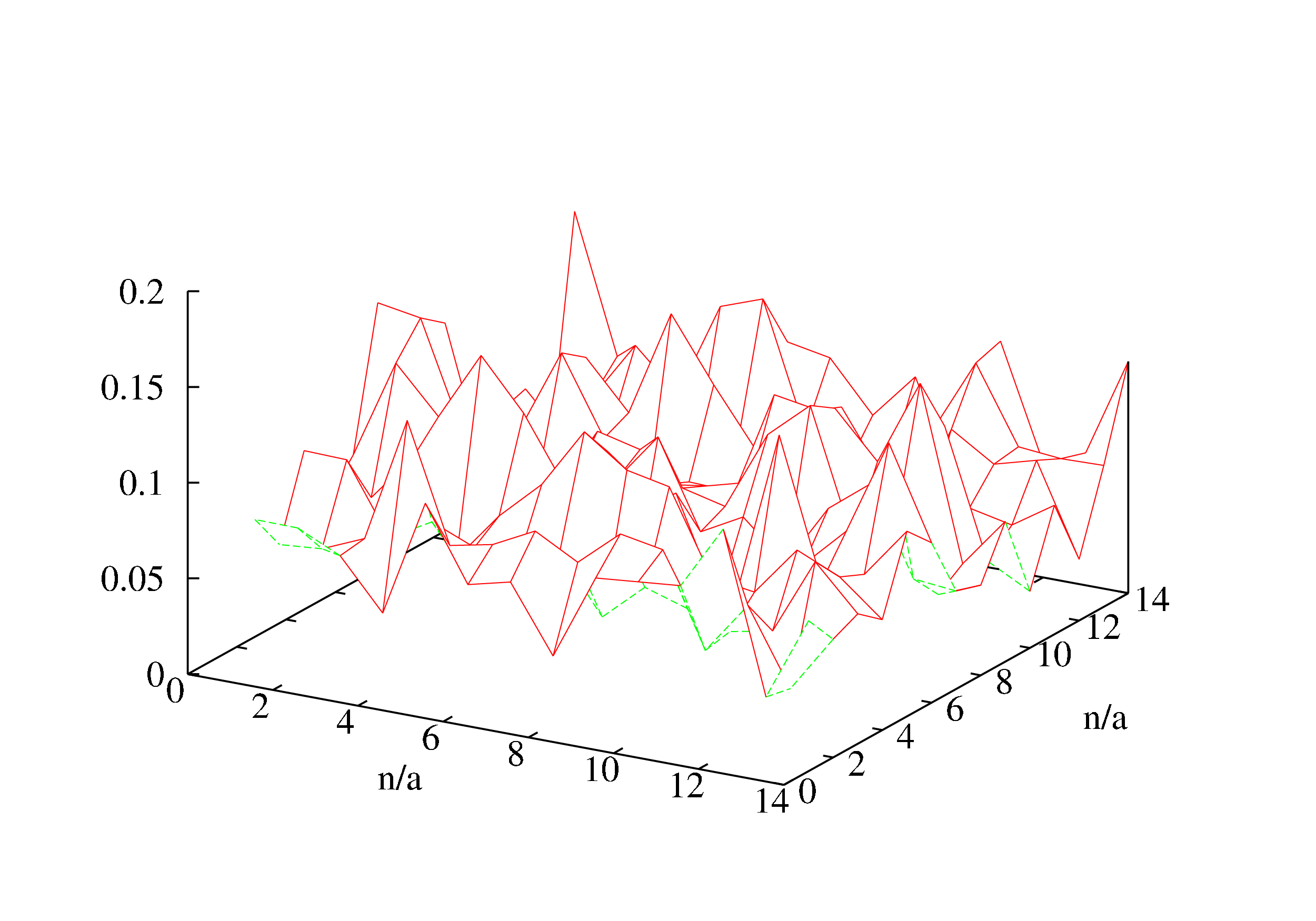}
\includegraphics[scale=0.25,angle=270]{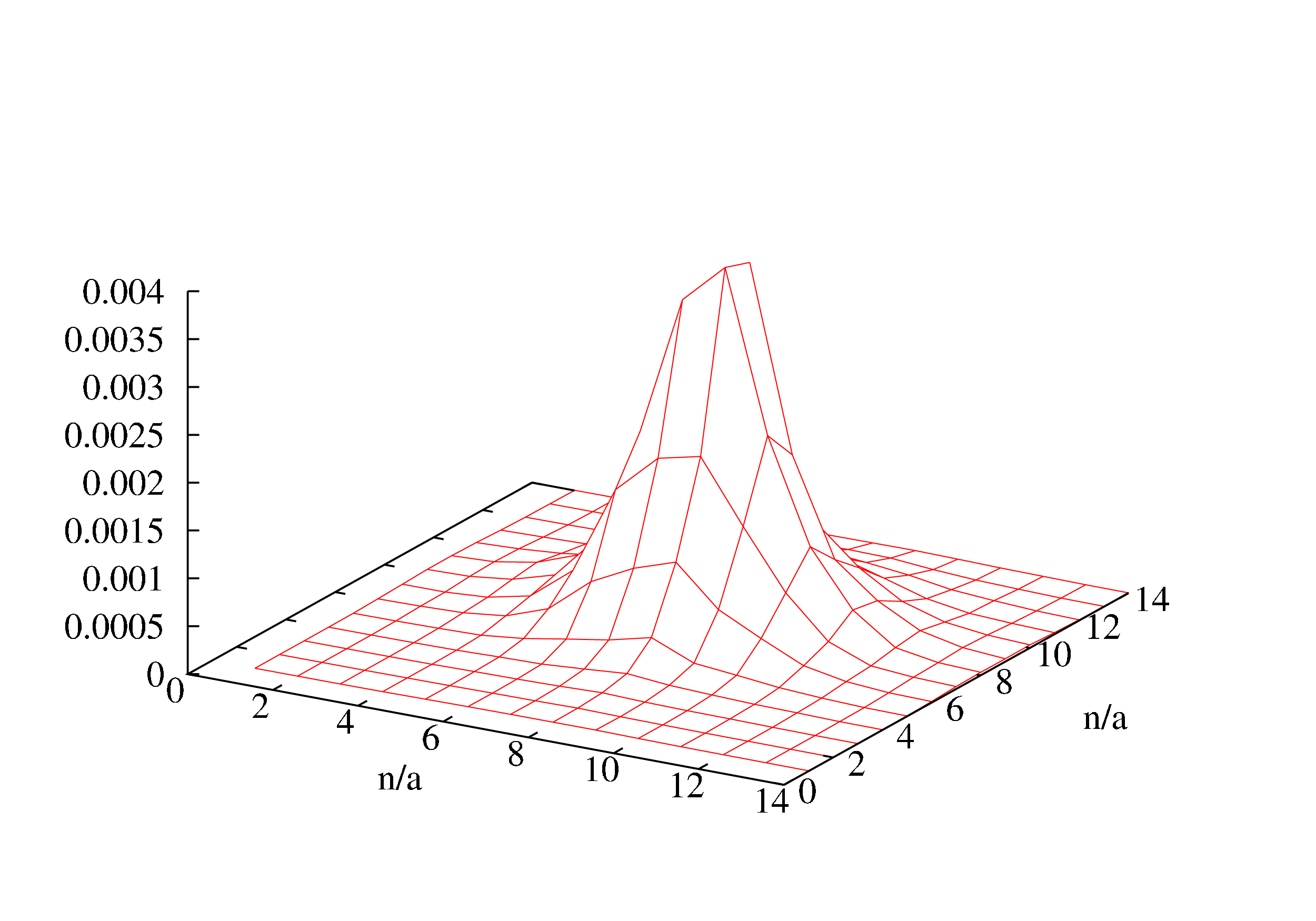}}
\caption{Comparison of the gauge action density (left)  and the \ssm\ density (right) 
 for a configuration generated by applying ten heat-bath sweeps ($\beta=4$)
 to a smooth instanton configuration.}
\label{fig:s_heat_4}
\end{figure}

The most important test, prior to its application to Monte Carlo
configurations, is to show that the procedure does indeed
filter out high frequency noise from the starting configuration, 
exhibiting its  long-range structures.  For this purpose, we 
began by generating  several smooth instanton configurations and  
added random noise to them. The way this was done was by applying to
the configuration a small number of heat bath updates corresponding to a 
Wilson action with different values of $\beta$. The small number of 
updates and the large values of $\beta$ guarantee that the instanton  
is not destroyed in the process, but considerable noise is added. 
The results presented here correspond to an initial instanton of size $\rho=3.4\,a$, on a periodic/twisted
lattice of size $14^4$, to which ten heat bath sweeps (with $\beta=30, 20, 8, 7, 6,5 $, and $4$) have been applied. 
The process is repeated with ten different initial random seeds giving
rise to a  heated instanton {\it ensemble} for each $\beta$ value and BC.
A characteristic example is shown in Fig.~\ref{fig:s_heat_4}, where we
display the \ssm\ and the gauge action densities for one rough configuration
with $\beta=4$ and PBC. While the action density dramatically roughens under heating, the
\ssm\ density is practically insensitive to the high frequency modes. The same is observed 
in the twisted case and for all the $\beta$ values analysed. This is a
good proof of the extraordinary filtering capacity of our method

\begin{figure}
\centering{
\includegraphics[scale=0.3,angle=270]{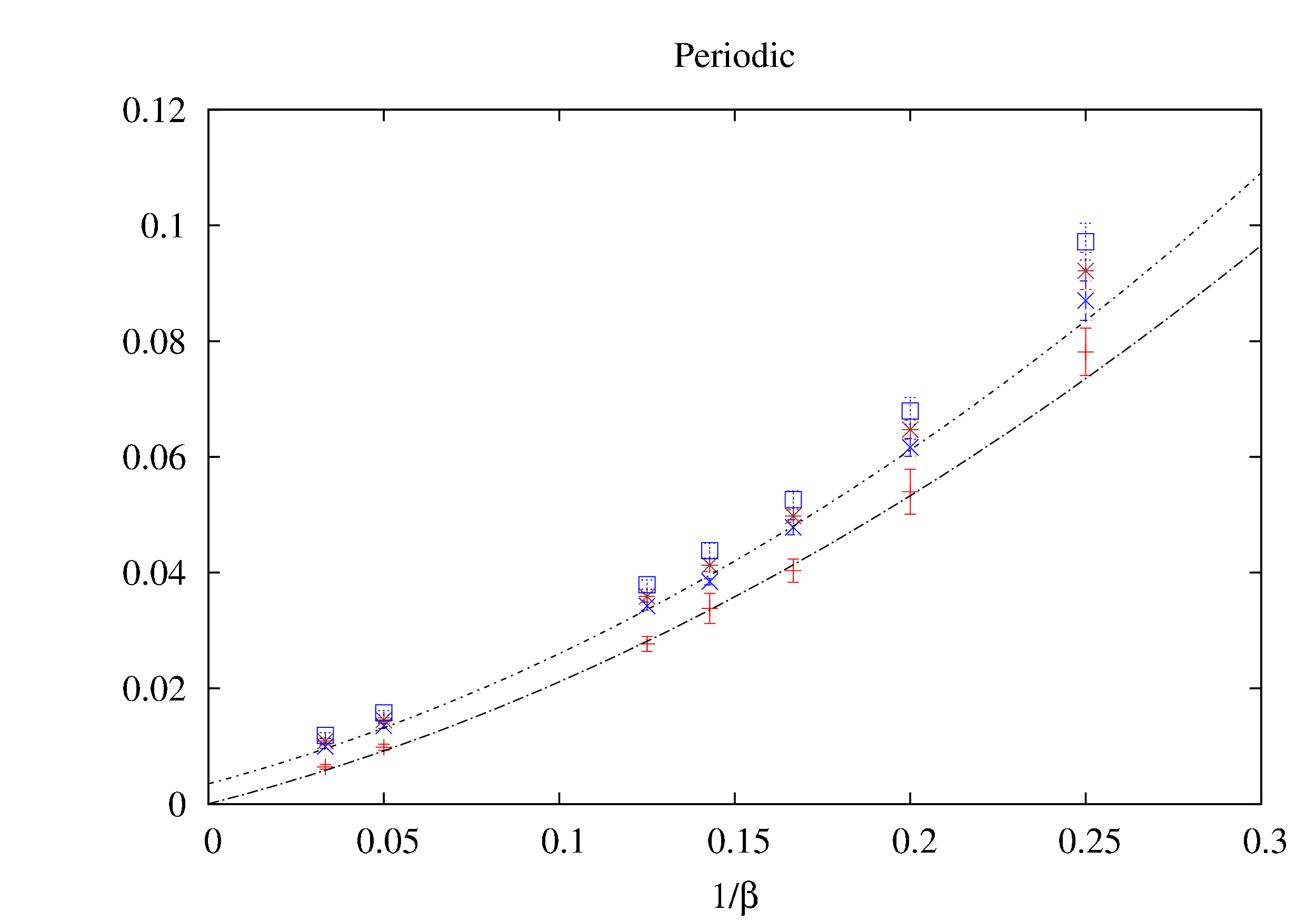}
\includegraphics[scale=0.3,angle=270]{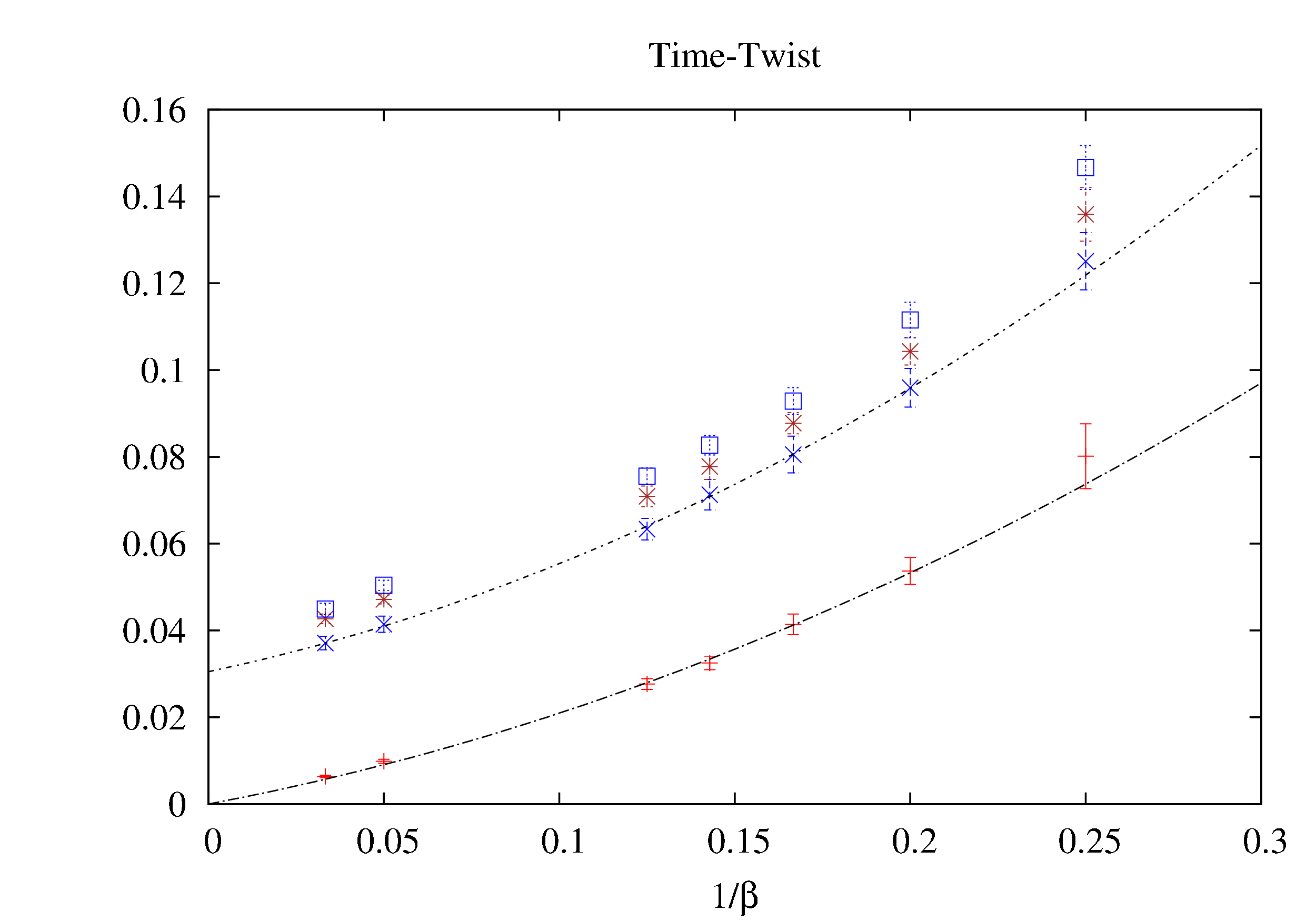}}
\caption{We display the four lowest eigenvalues of the $O^+$ operator for the ensemble of heated 
instanton configurations described in section 3.3.  For each value of $\beta$ we plot the eigenvalues averaged 
over the ensemble of heated configurations, with error bars corresponding to the variance of the sample.
The continuum lines correspond to a second order degree polynomial fit in $\beta^{-1}$.} 
\label{fig:heat_b}
\end{figure}

The ability of the method to recover the initial instanton structure
is related to the fact that, despite the eigenvalue becoming non-zero 
after the addition of noise, there seems to be no level crossing and
the \ssm\ is always cleanly separated. This is clearly seen in
Fig.~\ref{fig:heat_b}  where we display the four lowest eigenvalues
of the $O^+$ operator as a function of  $\beta^{-1}$, for periodic
(left), and twisted (right) boundary conditions. Notice that not only
is there no level crossing, but also the gap seems to remain constant
when the size of the noise (controlled by  $\beta^{-1}$) increases. 
The data for the lowest (\ssm) and first excited eigenvalues of
$O^\pm$ are well fitted by a second degree polynomial in
$\beta^{-1}$, which is the form predicted by perturbation theory
around the instanton configuration. The constant term is fixed to the 
value obtained for the instanton before heating (which for the ground
state is determined by finite-volume and finite lattice spacing
errors). The remaining coefficients turn to be very similar for both
eigenvalues and boundary conditions.

\section{Conclusions}
In this talk we have analysed the results of the filtering method
proposed in Ref.~\cite{Gonzalez-Arroyo2006} when applied to different
lattice configurations, some smooth and some rough. The results are
quite encouraging and point out to the main difficulties one might
encounter when applying the method to thermalised Monte Carlo 
configurations. Additional details and tests can be found in
Ref.~\cite{alfonsothesis,sso_mc}.

\end{document}